# Finite element method for thermal analysis of concentrating solar receivers

Stanko Shtrakov and Anton Stoilov
South-West University, Blagoevgrad, Bulgaria

Application of finite element method and heat conductivity transfer model for calculation of temperature distribution in receiver for dish-Stirling concentrating solar system is described. The method yields discretized equations that are entirely local to the elements and provides complete geometric flexibility. A computer program solving the finite element method problem is created and great number of numerical experiments is carried out. Illustrative numerical results are given for an array of triangular elements in receiver for dish-Stirling system.

1. **Introduction.**

Cavity receivers for solar concentrating systems absorb a concentrated solar energy, convert it to heat, and transfer heat to the working gas in power unit (Stirling engine, steam turbine). Heat flux in such elements is great and material tension is very high. Different shapes and materials were used to optimize the performance of cavity receivers.

Besides the natural experimental tests, there are possibilities to use serious theoretical studies for the thermal behavior of the cavity receivers. It would be very useful, because the processes in the cavity are complicated by high energy intensity and a real measure of temperature distribution is difficult and expensive.

Because of the special form of the solar receivers, it is not suitable to use ordinary mathematical technique, such as analytic methods or numeric solutions with finite difference approximation. Finite elements method is now one of the most popular ways to solve complicated mathematical problems, especially with irregular definition area. There are many works treating the heat conductivity problems by using finite elements methods and there is practical experience in this field.

In this paper is presented a mathematical model for heat conductivity processes in cavity receiver for dish-Stirling system with appliance of finite elements method for solving differential equations. The studies and numerical tests are made for a typical cylindrical cavity receiver, but algorithm and computer program created for calculation of temperature distribution in receiver are suited to solving a different forms of receivers.

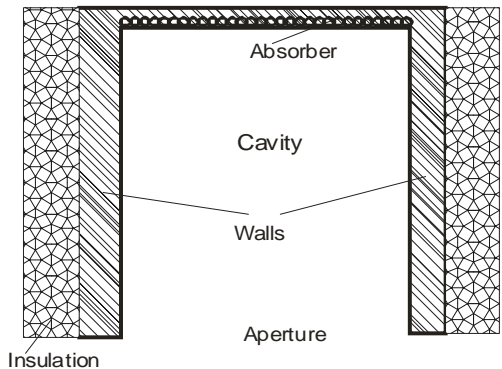

Fig. 1. Cavity receiver for dish-Stirling system

The simple construction scheme of typical tube type cavity receiver for dish-Stirling system is shown in Fig. 1. It comprises absorber plate (tube system for gas heating), cavity space, walls, aperture and insulation on the external surfaces of walls.

The absorbing surface is usually placed behind the focal point of the concentrator so that the flux density on the absorbing surface is reduced. The size of the absorber and cavity walls is typically kept to a minimum to reduce heat loss and receiver cost.

Concentrated radiation entering the receiver aperture diffuses inside the cavity. Most of the energy is directly absorbed by the absorber, and most of the remainder is reflected or reradiated within the cavity and is eventually absorbed by absorber and cavity walls. Converted by walls heat is conducted to the absorber plate and is transferred to the working gas. The major advantage of cavity receiver is that the size of the absorber may be different from the size of aperture. With a cavity receiver, the concentrator's focus is placed at the cavity aperture and the highly concentrated flux spread inside the cavity before encountering the larger absorbing surface area. This spreading reduces the flux incident on the absorber surface. When incident flux on the absorbing surface is high, it is difficult to transfer heat through surface without thermally overstressing materials.

The bottom of the receiver is a heat exchanger for the Stirling engine. Part of solar energy strikes directly the bottom wall of the receiver and by conduction it delivers heat to exchange pipes. Cylindrical walls absorb the other part of solar energy and conduct it to the bottom of the receiver. The external cylindrical surfaces are well insulated for protecting heat losses. Natural convection in cavity produces air circulation and convective losses. Heat losses are further caused by insensitive radiation from cavity aperture to the ambient. These losses are not big, because of the small measurements of the receiver. Nevertheless, all heat losses can be taking into account because they influence the temperature distribution in the receiver.

## 2. Theoretical Model

Mathematical model of processes in cavity receiver can be derived on the base of theoretical treating of heat conductivity in receiver walls, heat transfer to the working gas by absorber plate and heat losses to the ambient. Because of symmetry, the problem domain can be simply presented as in fig.2. The main mathematical equation in this problem is the heat conductivity equation for receiver construction written in cylindrical coordinate system:

$$\frac{\partial}{\partial r}\left(\lambda \frac{\partial T}{\partial r}\right) + \frac{\partial}{\partial z}\left(\lambda \frac{\partial T}{\partial z}\right) + \frac{1}{r}\lambda \frac{\partial T}{\partial r} = \Delta^2 T = 0 \qquad (1)$$

where **T** is the temperature in the walls, **r** and **z** – space variables in radial and high direction (fig.2) and **λ** – heat conductivity coefficient [W/m K]. $\nabla^2$ is Poason's operator for vector form description of equation.

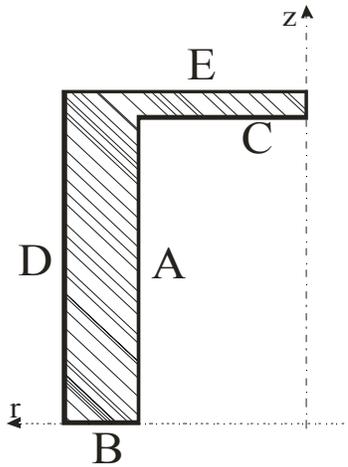

Fig.2 Theoretical scheme of receiver

The boundary conditions, needed for the solution of eqn (1), can be obtained from consideration of energy balance for different surfaces of the cavity receiver. External surfaces of the receiver (D) are insulated and the heat transfer to the ambient can be presented by the next equation:

$$\lambda \frac{\partial T}{\partial r} = K_D (T - T_a) \qquad (2)$$

Here $T_a$ is the ambient temperature, **T** – temperature of external surface and $K_d$ – overall heat transfer coefficient from the receiver external surface to the ambient air (including insulation).

The inner surfaces of receiver (A and C) exchange heat energy with air in cavity and absorb solar radiation entering the space in receiver. Boundary condition for these surfaces can be written as:

$$\lambda \frac{\partial T}{\partial r} = \alpha_B (T - T_f) + \overline{q} \qquad (3)$$

where $\alpha_B$ is a convective transfer coefficient in receiver cavity [W/m² K], $T_f$ – air temperature in cavity, **T** – temperature of inner receiver surface and **q** – solar energy flux [W/m²].

Surface **E** can be modeled as a heat exchanger for heating the working gas in Stirling motor. Boundary condition in this case is:

$$\lambda \frac{\partial T}{\partial r} = K_W (T - T_{wg}) \qquad (4)$$

Heat transfer coefficient $K_w$ depends on parameters of convective heat transfer to the working fluid. Temperature of working fluid $T_{wg}$ is defined from the thermodynamic treating of Stirling process. In general, boundary conditions can be written in common form as follow:

$$\lambda \frac{\partial T}{\partial n} = h^e T + C^e, \qquad (5)$$

where $h^e$ refers to the convection coefficient and $C^e$ describes the convection coefficient with the external temperature and solar flux. Superscript $e$ refers to the surface index ($e$). Derivative $\partial T/\partial n$ is taken to the outward normal direction $n$ for every boundary surface.

Differential problem (1) – (4) can be solved by numerical methods. The special form of problem domain (fig.2) leads to creating an unstructured grid for descretization of calculation area. This is the reason for using the finite element method for solving the problem.

### 3. Finite Element Method

In the finite element method (FEM), the problem domain is discretized and represented by an assembly of finite elements. The method yields discretized equations that are entirely local to the element. As a result, the discrete equations are developed in isolation and are independent of the mesh configuration. In this way, finite elements can readily accommodate unstructured and complex grids, unlike other numerical methods (finite difference method) requiring structured (rows and columns) format. Finite element method provides complete geometric flexibility.

Various approaches to the finite element method formulation are used, but the most prevalent is the Galerkin method [1,2]. It provides the correct number of basis function and the same resulting equations as methods based on other (more complicated) types of formulations.

Galerkin's method selects the weight functions equal to the basis functions (shape functions) of the approximate solution. It will be demonstrated on the example of solving the two-dimensional thermal conductivity problem in cavity receiver for dish-Stirling system – equation (1) with boundary condition (2) – (4).

Commonly encountered elements in two-dimensional configuration include a linear triangle, bilinear rectangle, and bilinear quadrilateral. Triangular elements are well suited to irregular boundaries, and spatial and scalar interpretation can be accommodated in terms of a linear polynomial.

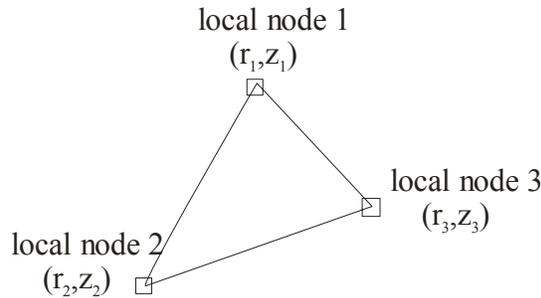

Fig. 3. Linear triangle

Nodes are assigned to location in the element at which unknown function, such as temperature, is to be determined. Nodes are often placed only at the corners of the elements (Fig.3), but additional nodes can be placed internally or along the element boundaries.

The main problem in finite elements method is to choose shape or integration function to provide approximate variation of the depended variable within each element between the values of the nodes. The simplest shape function is linear. Each shape function is a local interpretation function that is defined only within elements containing a particular node. Linear triangle element (Fig.3) refers to linear interpolation along a side or within the element. For an interpolation involving a scalar shape form function, $N(r,z)$ within a linear triangle can be described as:

$$N(r,z)=a+br+cz. \qquad (6)$$

where the unknown coefficients, $a, b, c,$ can be determined on the basis of the substitution of nodal values. For example, at node 1, the position is ($r_1, z_1$) and the scalar shape function is $N_1$.

By using the shape functions, an approximate solution $\hat{T}(r,z)$ for $T(r,z)$ is assumed in the form

$$\hat{T}(r,z) = \sum_{j=1}^{N} T_j N_j(r,z) \qquad (7)$$

where $T_j$ are the temperatures at the nodes desired for the solution, and $N_j$ are the shape functions that are equal to *1* at each node. When this approximation is substituted into the energy equation (1), there is a residual that depends on *r* and *z*:

$$\frac{\partial}{\partial r}\left(\lambda \frac{\partial \hat{T}}{\partial r}\right) + \frac{\partial}{\partial z}\left(\lambda \frac{\partial \hat{T}}{\partial z}\right) + \frac{1}{r}\lambda \frac{\partial \hat{T}}{\partial r} = \Delta^2 \hat{T} = \operatorname{Re} s(r,z) \qquad (8)$$

It is desired to be obtained a solution that, in average sense over the entire volume, is as close as possible to the exact solution each time. Variational principles are applied to minimize the residual. A set $W_i(r,z)$ of independent weighting function is applied, and the residual is made orthogonal with respect to each of the weighting functions. This provides the following integral, which is evaluated for each of the sets of independent weighting functions,

$$\iint_S \operatorname{Re} s(r,z) W_i(r,z) dS \qquad (9)$$

The integration is over the whole volume in which the solution is being obtained. In the Galerkin method the weighting functions are chosen to be the same function set as the shape functions. Equation (9) provides the Galerkin form of energy equation for each of the weighting functions,

$$\iint_S \left[\lambda \cdot \sum_{j=1}^{N} T_j \nabla^2 N_j(r,z)\right] \cdot N_i(r,z) \cdot dS \qquad (10)$$

Boundary conditions can be incorporated through the boundary integral term in above equation. Considering the general form of the boundary conditions (5) and using integration by parts, the following form of equation (10) can be received [1]:

$$\iint_S \left[\lambda \sum_{j=1}^{N} \left(\frac{\partial N_i}{\partial r}\frac{\partial N_j}{\partial r} + \frac{\partial N_i}{\partial z}\frac{\partial N_j}{\partial z} + \frac{1}{r}\frac{\partial N_j}{\partial r} N_i\right) T_j\right] \cdot dS = \int_n [N_i(r,z)(h^e T_i + C^e)] dn \qquad (11)$$

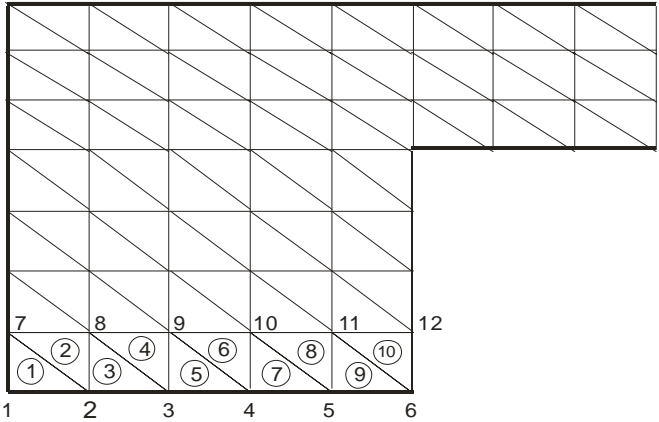

Fig. 4. Discretization of domain

Evaluating equation (11) for each *i* provides *N* simultaneous algebraic equations for the $\hat{T}(r,z)$. In the matrix form, this is:

$$[K_{ij}] [T_j] = [F_j] \qquad (12)$$

Since each shape function $N_j(r,z)$ (and hence each weighting function in Galerkin method) is zero except within an element containing $T_j$, the resulting matrix $[K_{ij}]$ for solving the simultaneous equations for $T_j$ is banded and sparse. It is usually banded along the diagonal (banded matrix). The finite element method yields discretized equations that are entirely local to the element and hence the global matrix $[K_{ij}]$ is a simple combination (sum) from local matrixes of each node.

Solving the system of algebraic equations (12), the unknown temperatures $T_j$ can be received as a result. Because of the special form of $[K_{ij}]$ matrix, different numerical techniques have been developed [1,2] for solving the system (12).

### 4. Numerical solution techniques

The mentioned above technique is demonstrated on the example of cavity receiver (fig.1). Using triangle elements the problem domain can be discretized as it is shown in fig.4. Each element is numbered and nodes coordinates $(r_i, z_i)$ are specified.

Interpolation of the temperature in triangle element can be written in terms of the shape functions, $N_i$, $N_j$, $N_k$, as follow:

$T^{(e)}(r,z) = T_i N_i^{(e)}(r,z) + T_j N_j^{(e)}(r,z) + T_k N_k^{(e)}(r,z)$ where function $N_i^e = a_i^e + b_i^e r + c_i^e z$ ($i$ =1, 2, 3) is determined by considering the next conditions: $N_i^e(r_i, z_i) = 1 \quad N_i^e(r_j, z_j) = N_i^e(r_k, z_k) = 0$.

Superscription $e$ refers to the element number ($e$). Coefficients $a$, $b$, and $c$ in shape form functions have the following form:

$$\begin{cases} a_i = (r_j z_k - r_k z_j)/2S \\ b_i = (z_j - z_k)/2S \\ c_i = (r_k - r_j)/2S \end{cases} \quad \begin{cases} a_j = (r_k z_i - r_k z_i)/2S \\ b_j = (z_k - z_i)/2S \\ c_j = (r_i - z_k)/2S \end{cases} \quad \begin{cases} a_k = (r_i z_j - r_j z_i)/2S \\ b_k = (z_i - z_j)/2S \\ c_k = (r_j - r_i)/2S \end{cases} \quad (13)$$

where S is the area of triangle element.

Finite element method is well-established numerical technique for heat conduction problem in orthogonal coordinate system. Heat conductive problem in cylindrical coordinate system (equation (1)) differs from the orthogonal coordinate problem with next integral in equation (11):

$$\lambda \iint_S \frac{1}{r} \frac{\partial N_j}{\partial r} N_i \, dr \, d \qquad (14)$$

The shape function derivatives in equation (11) are constants: $\partial N_i/\partial r = b_i$ and $\partial N_i/\partial z = b_i$. This gives a simple form of the local matrix for element $n$ (part of global matrix $[K_{ij}]$) in orthogonal coordinate system (without contribution of integral (14)). With regarding the above values of derivatives and not considering the boundary conditions the matrix can be written in the following form [1]:

$$k^{(n)} = \lambda \begin{bmatrix} (b_i^n)^2 + (c_i^n)^2 & b_i^n b_j^n + c_i^n c_j^n & b_i^n b_k^n + c_i^n c_k^n \\ b_i^n b_j^n + c_i^n c_j^n & (b_j^n)^2 + (c_j^n)^2 & b_j^n b_k^n + c_j^n c_k^n \\ b_i^n b_k^n + c_i^n c_k^n & b_j^n b_k^n + c_j^n c_k^n & (b_k^n)^2 + (c_k^n)^2 \end{bmatrix} \qquad (15)$$

Integral (14) is more complicated because there is $1/r$ member and $N_i$ is in form (6) with coordinates $r$ and $z$. The integral (14) can be divided in three parts:

$$I = \lambda \cdot b_j \iint \frac{N_i}{r} dr dz = \lambda \cdot b_j \left[ \iint \frac{a_i}{r} dr dz + \iint b_i dr dz + \iint c_i \frac{z}{r} dr dz \right] = \lambda \cdot b_j [I_1 + I_2 + I_3]$$

Integral $I_2$ is easy to be calculated because $b_i$ is constant and: $I_2 = b_i S$, where S is the area of triangle element. Other integrals can be calculated by integrating over triangle area (fig. 5). By using the triangle elements with particular disposition as it is shown in fig.4 and dimensions presented on the fig.5 the integrals $I_1$ and $I_3$ can be presented as:

$$I_1 = \frac{\alpha_i r_2}{\mu} [1 + v(\ln v - 1)]$$

$$I_3 = \frac{c_i}{\mu} [r_1 \ln v (z_2 + \frac{r_1}{2\mu}) - z_2 \nabla r (r_1 r_2 - \frac{3}{4} r_1^2 - \frac{r_1^2}{4})] \qquad (17)$$

where $\mu = \dfrac{\nabla r}{\nabla z}$ и $\nu = \dfrac{r_2}{r_1}$

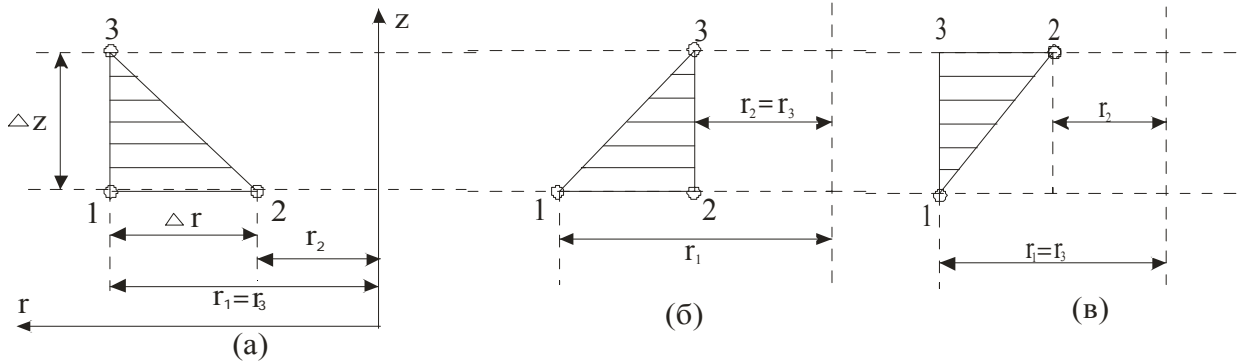

Fig.5 Triangle elements

These values must be added to the corresponding members in matrix *[K$_{ij}$]*.

The cylindrical heat conductivity problem can be solved by approximate methods. It is necessary in case of common disposition and form of triangle elements. One of these methods is to make integration in (14) assuming coordinates *r* and *z* as constants and equal to the coordinates of the mass center of the element - *r$_m$* and *z$_m$*. In this case approximate solution can be received, and it will be so close to the real solution as much as small are elements. Integral (16) can be presented as:

$$I = b_j S^e \left[ \dfrac{a_i}{r_m} + b_i + c_i \dfrac{z_m}{r_m} \right] \qquad (18)$$

Other approximate method is determined by rearranging the original equation (1) in form:

$$\dfrac{\partial}{\partial r}\left( \lambda r \dfrac{\partial T}{\partial r} \right) + \dfrac{\partial}{\partial z}\left( \lambda r \dfrac{\partial T}{\partial z} \right) = 0 \qquad (19)$$

This form resembles to the form of conductivity equation written in Decart's coordinates if product *λr* is interpreted as modified conductivity coefficient. It will vary for different elements and will depend on radius *r*. For integration in (11) it can be assumed approximate approach, for example to use constant radius *r$_m$* of the mass center of element. In this case the matrix (15) can be used, but the coefficient λ will be different for each element of the matrix.

Boundary conditions (5) contribute the matrix elements by adding values for triangle sides that touch the boundaries of the receiver: surfaces A, B, C, D, E (fig.2). Contribution is determined by integrating the last members in (11). It differs for elements with equal indexes and with different indexes:

*g$_{ii}$* = *g$_{jj}$* = *g$_{kk}$* = *h$^e$*Δr/3 ; *g$_{ij}$* = *g$_{jk}$* = *g$_{ik}$* = *h$^e$*Δr/6 ; (20)

Values in right column of matrix equation are determined for triangle sides that contact the ambient or exchange elements of receiver. These are formed by integrating the members not containing temperature in last part of (11):

$$F_i = \dfrac{1}{2} \cdot C^e \cdot \nabla r \quad \text{or} \quad F_i = \dfrac{1}{2} \cdot C^e \cdot \nabla z \qquad (21)$$

**5. Programming and result**

A computer program solving the finite element problem has been created and great number of numerical experiments has been carried out. The program comprises two main parts. The first part includes program modules for automatic discretization and numbering the elements and nodes in mesh configuration. These modules are specific for different problems and configuration of domain. The main result of performance of program modules is forming the matrix with numbers and coordinates of triangle element nodes.

The second part of the program modules uses created in the first module data to form the basic matrix $K_{ij}$ – equation (12). Special procedure for solving the band matrix is used to calculate the requested parameters (temperatures, heat flux etc.). This part is standard and can be used for solving different tasks.

The program and mathematical treating of finite element method were verified by using the exact analytical solution of thermal conductivity problem. It is known, that for simple configuration of domain, the heat conductivity problem has analytical solution. Such configuration is a cylindrical domain. For such a simple configuration the finite element method gives results, which a very close to the analytical solution (difference is below 0.1%).

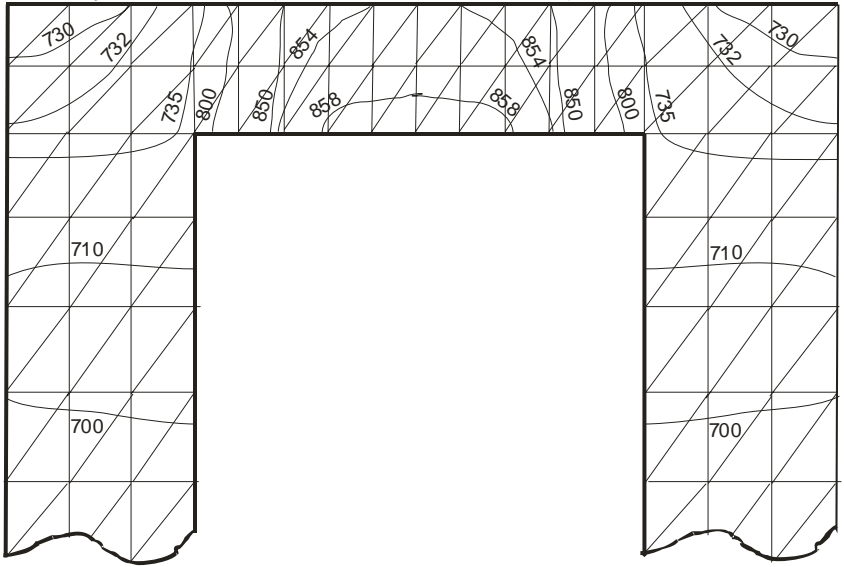

Fig.7 Temperature distribution in Receiver

Fig. 7 presents the temperature distribution in receiver for dish-Stirling system, calculated by computer program. Some of the used data for boundary conditions are assessed approximately. This means that the temperature distribution in receiver must be interpreted only in rough figures. When the conditions for receiver performance are defined more precise, the results for temperature distribution will be more accurately.

**6. Conclusion**

Finite element method has been applied for heat transfer problem in receiver for dish-Stirling system. It uses unstructured and complex grids, which are suited to the special forms of absorber element. Finite element method provides complete geometric flexibility.

Presented mathematical model and algorithm for program is universal and can be used for different tasks. In this paper are discussed only principle aspects of using the finite element method for heat transfer processes in receiver for solar receivers in concentrating solar systems. Numerical experiments carried out in this research show that this technique gives very good approaches to the real thermal processes. It can be used successfully to investigate different conditions and forms of receivers for concentrating solar systems and other thermal equipments.